\documentclass[a4paper,scale=0.8,onecolumn,nobibnotes,nofootinbib]{revtex4}

\usepackage{amsmath,amssymb}
\usepackage{graphicx,subfigure,epsfig}
\usepackage{hyperref}
\usepackage{geometry}
\hypersetup{
    colorlinks=true,
    citecolor=blue,
    linkcolor=red,
    filecolor=magenta,
    urlcolor=cyan,}

\newcommand{\be}{\begin{equation}}
\newcommand{\ee}{\end{equation}}

\begin{document}

\title{Research on the new form of higher-order generalized uncertainty principle \\ in quantum system}
\author{Zi-Long Zhao$^1$, Qi-Kang Ran $^2$, Hassan Hassanabadi$^{3,}$, \\ Yi Yang$^{1,}$, Hao Chen$^{1,}$, Zheng-Wen Long$^{1,}$\footnote{Corresponding author.\\
\textsl{E-mail address:} zwlong@gzu.edu.cn}}
\affiliation{$^1$ Department of Physics, Guizhou University, Guiyang, 550025, China\\
$^2$ College of Mathematics, Shanghai University of Finance and Economics, Shanghai 200433, China\\
$^3$ Faculty of Physics, Shahrood University of Technology, Shahrood, Iran}

\date{\today}

\begin{abstract}
This paper proposes a new high-order generalized uncertainty principle, which can modify the momentum operator and position operator simultaneously. Moreover, the new form of GUP is consistent with the viewpoint of the existence of the minimum length uncertainty and the maximum observable momentum proposed by the mainstream quantum gravity theory. By using the new GUP, the maximum localization state and position eigenfunction are discussed, and the corresponding conclusions are compared with the existing literature. The harmonic oscillator is further discussed at the end of this article as an example.\\

\end{abstract}
\maketitle
~~~~~~~~~
\section{Introduction}
Relativity and quantum theory are the most epoch-making significant progress in physics in the 20th century, whose generation and development can be said to have run through the development of physics in the 20th century. It is no exaggeration to say that the development of physics in the 20th century is basically along the two main lines of relativity and quantum theory. Relativity can reasonably explain the physical phenomena in the large-scale cosmic field, and its predicted results have been well verified in the solar system and some binary pulsar systems \cite{ref a1}, while the quantum mechanics and its offspring quantum field theory are in good agreement with the experimental results in the micro field, and its theoretical prediction and experimental consistency is incredible. However, although these two theories have achieved great experimental successes in their respective fields, they are also faced with a series of acute problems, especially for the case where the mass of matter is extremely large and the scale is extremely small, any one of the two theories cannot make a good explanation. The theoretical root lies in the incompatibility between the quantized description of matter and the geometric description of space-time. Therefore, establishing an over-arching theoretical framework can accommodate both relativity and quantum mechanics, which has been a goal pursued by many physicists for a long time \cite{ref a3}.

Quantum gravity, as a general name for many developmental candidate unified theories that is getting more and more recognized in physics, aims to combine quantum mechanics with general relativity, but it is not the only determined theory \cite{ref a3,ref a4,ref1,ref2,ref3,ref4,ref5,ref7}. At present, the mainstream theories of quantum gravity in physics mainly include loop quantum gravity \cite{ref5}, string theory \cite{ref8,ref9,ref10,ref11}, non-commutative space-time \cite{ref12,ref13,ref14,ref15}, doubly special relativity (DSR) \cite{ref16,ref17,ref18,ref19} and black holes Gedanken experiments \cite{ref20,ref21}, etc., which all point out that there is a non-zero minimal observable length of Planck scale that is about $10^{-35}$ m. However, in the field of quantum gravity, an unavoidable problem is the existence of curvature of space-time. Especially in the case of focusing large energy in a small volume, when the energy density increases to a certain extent, the effect of space-time curvature will become more and more significant, and if the energy density exceeds the Planck energy scale, the effect of gravity is so strong that it will inevitably leads to discreteness of the space-time, which naturally introduce the concept of minimal observable length. What is exciting is that the normalization of quantum field theory under a curved background can be achieved by introducing a minimum observable length as a natural cutoff in the ultraviolet region, which avoids the usual ultraviolet divergence \cite{ref7,ref22}. In addition, the concept of minimum measured length also appears in string theory. Due to the string cannot probe the distance smaller than the string scale in string theory, and the string excitations may occur and lead to its own extension if the energy of a string approaches the Planck scale \cite{ref7,ref9,ref12,ref22}.

However, according to the Heisenberg uncertainty principle of standard quantum mechanics, $\Delta x\Delta p\geq\hbar/2$, the minimal uncertainty length will gradually tends to be zero as momentum approach to infinity, which does not give a reasonable explanation for this fact of exist a minimal measurable length \cite{ref23}. Therefore, it is necessary to introduce the concept of minimal length into quantum mechanics by modifying the Heisenberg uncertainty principle, which is called the generalized uncertainty principle (GUP). In the past two decades, the influence of GUP in quantum mechanical systems has been widely discussed in many articles as a hot topic \cite{ref4,ref5,ref7,ref22,ref23,ref b1,ref b2,ref b3,ref b4,ref b5,ref b6,ref b7,ref b8,ref b9,ref b10,ref b11,ref b12,ref b13,ref b14,ref b15,ref b16,ref b17,ref b18}. However, especially in the discussion of problems in the context of quantum gravity, the curvature of space-time will become an unavoidable topic. As we know that the momentum space in the curved space-time is also curved, and there is no coordinate transformation that can make the curvature of momentum space disappearing \cite{ref4}. Moreover, in recent literatures \cite{ref4,ref24,ref25,ref26,ref27,ref28}, many scholars have given conclusions consistent with the DSR by studying the construction of the perturbation GUP, which suggests that the minimal uncertainty of the momentum may come from the curvature of space-time \cite{ref22,ref29}. Therefore, it is necessary to propose a new form of GUP to satisfy the features of curved space-time.

The structure of this article is as follows: the new form of the modification of Heisenberg uncertainty principle in the momentum representation is discussed in detail in section 2. In section 3, we studied the functional analysis of the position operator. And the maximal localization states were discussed in section 4. In section 5, we take harmonic oscillator as an example to carry out relevant research. In the last part, the conclusion was contributed.
\section{Review of the generalized uncertainty principle and construct the new one}
Now, let's first review the development of generalized uncertainty principle. As we all know, GUP was originally introduced by Kempf, Mangano and Mann (KMM) in their seminal paper, and its deformed commutation relation was written as \cite{ref13}
\begin{equation}\label{eq:1}
\;\;\;\;\;\;\;\;\;\left[ {X,P} \right] = i\hbar \left( {1 + \beta {P^2}} \right).
\end{equation}
So, the uncertainty relation satisfies
\begin{equation}\label{eq:2}
\;\;\;\;\;\;\;\;\;\left( {\Delta X} \right)\left( {\Delta P} \right) \ge \frac{\hbar }{2}\left[ {1 + \beta {{\left( {\Delta P} \right)}^2}} \right],
\end{equation}
in which indicates that there is a fundamental minimal length scale as following
\begin{equation}\label{eq:3}
\;\;\;\;\;\;\;\;\;\left( {\Delta X} \right)_{\min }^{{\rm{KMM}}} = \hbar \sqrt \beta.
\end{equation}
Here, the parameters assume that $\beta=\beta_0/{M_{pl}}{c^2}{\sim}{10^{19}}\\{\rm{GeV}}$,
where $M_{pl}$ is the Planck mass and $\beta_0$ is of the order of unity. However, although Kempf et al. have successfully introduced the concept of minimum length into GUP and it has also been used to obtain a series of conclusions in the subsequent research, the flaw is that Eq. (\ref{eq:1}) could not derive the concept of maximum momentum. Based on the work of the predecessors, Ali, Das and Vagenas further introduced the idea of maximum momentum, and proposed the modified commutation relation as follows \cite{ref25,ref26,ref27}
\begin{eqnarray}\label{eq:4}
\;\;\;\;\;\;\;\;\;\left[ {{X_i},\;{P_i}} \right] = i\hbar \left[ {{\delta _{ij}} - \alpha \left( {P{\delta _{ij}} + \frac{{{P_i}{P_j}}}{P}} \right)} \right. \left. { + {\alpha ^2}\left( {{P^2}{\delta _{ij}} + 3{P_i}{P_j}} \right)} \right],
\end{eqnarray}
in which ${p^2} = \sum\nolimits_{j = 1}^3 {{P_i}{P_j}}$, $\alpha=\alpha_0/{M_{pl}}{c}={\alpha_0}{l_{pl}}/\hbar$, $l_{pl}\approx {10^{-35}}\;\rm m =\rm{Planck\;length}$, $M_{pl}=\rm{Planck\;mass}$, and $M_{pl}{c^2}=\rm{Planck\;energy}\approx{10^{19}}\rm{GeV}$. Thus, according to the Eq. (\ref{eq:4}) the maximal observable momentum and the minimal observable length could be written as
\begin{eqnarray}\label{eq:5}
\;\;\;\;\;\;\;\;\;\Delta X \ge \left( {\Delta X} \right)_{\min }^{{\rm{ADV}}} \approx {\alpha _0}{l_{pl}} = \hbar \alpha, 
\;\;\;\;\Delta P \le \left( {\Delta P} \right)_{\max }^{{\rm{ADV}}} \approx {{{M_{pl}}c} \mathord{\left/  {\vphantom {{{M_{pl}}c} {{\alpha _0}}}} \right.  \kern-\nulldelimiterspace} {{\alpha _0}}} = {1 \mathord{\left/  {\vphantom {1 \alpha }} \right.  \kern-\nulldelimiterspace} \alpha}.
\end{eqnarray}
Similarly, the momentum and position operators satisfying the commutative relation (\ref{eq:4}) could also be given
\begin{equation}\label{eq:6}
\;\;\;\;\;\;\;\;\;\;\;\;X_{i}=x_{i},\;\;\;\;\;P_{i}=p_{i}({1-\alpha p+2{\alpha}^2{p^2}}),
\end{equation}
in which $x_{i}$ and $p_{i}$ agree with the canonical commutation relations $[x_{i},p_{j}]=i\hbar\delta_{ij}$ and $p = \sqrt {\sum\nolimits_{j = 1}^3 {{p_j}{p_j}} }$. Although Ali et al. have successfully introduced the concept of maximal momentum and minimal length into GUP, but this proposal also has some difficulties in itself. Since it considers the perturbation effect itself, that is to say, this proposal is only valid when the GUP parameter is very small \cite{ref25,ref30}. Moreover, the maximal momentum uncertainty is not equivalent to the concept of the viewpoint of maximal momentum in DSR. In fact, the Eq. (\ref{eq:5}) gives only the upper limit of the uncertainty of momentum measurement, rather than the value of observed momentum. Thus, as a long-term hot topic, in recent years, Pedram has proposed a new form of higher order GUP in order to deal with the difficulties in the above modifications \cite{ref22,ref31}
\begin{equation}\label{eq:7}
\;\;\;\;\;\;\;\;\;\;\;\;\left[ {X,\;P} \right] = \frac{{i\hbar }}{{1 - \beta {p^2}}}.
\end{equation}
It is not difficult to find that the commutative relation proposed by Pedram not only fully agrees with Eq. (\ref{eq:1}) to the leading order, but also seems to be more meaningful. Since Eq. (\ref{eq:7}) contains a singularity at $p^2=1/\beta$, it shows that the momentum of a particle cannot exceed $1/\sqrt\beta$, which is completely consistent with the conclusion of DSR \cite{ref16,ref17,ref18,ref19}. Furthermore, the uncertainty relation that was produced by Eq. (\ref{eq:7}) can be written as
\begin{eqnarray}\label{eq:8}
&&\left( {\Delta X} \right)\left( {\Delta P} \right) \ge \frac{\hbar }{2}\left\langle {\frac{1}{{1 - \beta {p^2}}}} \right\rangle \nonumber \\
&&\;\;\;\ge \frac{\hbar }{2}\left( {1 + \beta \left\langle {{p^2}} \right\rangle  + {\beta ^2}\left\langle {{p^4}} \right\rangle  + {\beta ^3}\left\langle {{p^6}} \right\rangle  +  \cdot  \cdot  \cdot } \right)\nonumber \\
&&\;\;\;\ge \frac{\hbar }{2}\left\{{1 + \beta \left[ {{{\left( {\Delta p} \right)}^2} + {{\left\langle p \right\rangle }^2}} \right] + {\beta ^2}{{\left[ {{{\left( {\Delta p} \right)}^2} + {{\left\langle p \right\rangle }^2}} \right]}^2}} \right.\\
&&\;\;\;\;\;\;\left. { + {\beta ^3}{{\left[ {{{\left( {\Delta p} \right)}^2} + {{\left\langle p \right\rangle }^2}} \right]}^3} +  \cdot  \cdot  \cdot } \right\}\nonumber \\
&&\;\;\; \ge \frac{{{\hbar  \mathord{\left/ {\vphantom {\hbar  2}} \right.  \kern-\nulldelimiterspace} 2}}}{{1 - \beta \left[ {{{\left( {\Delta p} \right)}^2} - {{\left\langle p \right\rangle }^2}} \right]}}\nonumber.
\end{eqnarray}
Here, we assume $\langle {p^2}\rangle  = {\langle p \rangle ^2} + {\left( {\Delta p} \right)^2}$ and use the property $ \langle {p^{2n}}\rangle  \ge {\langle {p^2}\rangle ^n}$ \cite{ref32}. The minimal length uncertainty of deformation algebra (\ref{eq:8}) can only be reached for $\langle p\rangle=0$. Thus, the Eq. (\ref{eq:8}) could be rewrite as following
\begin{equation}\label{eq:9}
\;\;\;\;\;\;\;\;\;\;\;\;\left( {\Delta X} \right)\left( {\Delta P} \right) = \frac{{{\hbar  \mathord{\left/ {\vphantom {\hbar  2}} \right.  \kern-\nulldelimiterspace} 2}}}{{1 - \beta {{\left( {\Delta p} \right)}^2}}},
\end{equation}
in which implies there is a minimal value at ${\Delta p}=1/{\sqrt{3\beta}}$. And the absolute minimal uncertainty of position could be given
\begin{equation}\label{eq:10}
\;\;\;\;\;\;\;\;\;\;\;\;\left( {\Delta X} \right)_{\min }^{{\rm{PP}}} = \frac{{3\sqrt 3 }}{4}\hbar \sqrt \beta.
\end{equation}
Through the above review, it is easy to find that the GUP with different algebraic structure will give the new implications to the Hilbert space representation in quantum mechanics. Therefore, in this paper, we will focus on constructing a new form of generalized uncertainty principle to satisfy the feature of curved space-time. This is because the curvature as the basic feature of curved space-time will become an unavoidable topic, especially when related issues involve quantum gravity. As we know, the momentum space in curved space-time is also curved, and no coordinate transformation can make the curvature of momentum space disappearing \cite{ref4}. In addition, there are some scholars have put forward the viewpoint in their researches that the minimal uncertainty of momentum may come from the curvature of space-time \cite{ref29}. Therefore, it is meaningful and necessary to take the curvature of the curved space-time into the generalized uncertainty principle. So, the new forms of momentum and coordinate operators are constructed as follows
\begin{eqnarray}\label{eq:11}
\;\;\;\;\;\;\;\;\;\;\;P \cdot \psi \left( p \right) = \frac{p}{{\sqrt {1 - \beta {p^2}} }} \cdot \psi \left( p \right), 
\;\;\;\;\;X \cdot \psi \left( p \right) = i\hbar \sqrt {1 - \beta {p^2}} \frac{d}{{dp}}\psi \left( p \right).
\end{eqnarray}
It is easy to verify that the new form of operators were constructed in momentum space of this paper not only satisfy commutation relationship proposed by Pedram, but also seem to have more abundant physical connotation. This is mainly reflected in this fact that the basic feature of space-time was taken into account in the modification of Heisenberg uncertainty relation. Generally speaking, the main difference is that this paper does not modify one of the momentum or the position operators as other literatures \cite{ref13,ref19,ref30,ref31}, but also modify both simultaneously. The main reason why this modification is adopted is that the momentum space of curved space-time is also curved, and there is no position transformation to make curvature disappear. In addition, it can also be see that the new form of GUP was constructed in this paper not only pointed out the existence of minimal observation length, but also pointed out that the momentum of particles cannot exceed $1/\sqrt\beta$, which fully agrees with the main conclusions of previous studies \cite{ref16,ref17,ref18,ref19,ref30,ref31}. Therefore, the new GUP can be regarded as an attempt to consider the curvature effect in the modified Heisenberg uncertainty principle.

And it is not difficult to see that in the dense domain ${S_\infty }$ the function decays faster than any power. Moreover, we can see that the position operator and momentum operator are symmetric in the domain ${S_\infty }$
\begin{eqnarray}\label{eq:12}
\;\;\;\;\;\;\;\;\;\;\left( {\left\langle \psi  \right|P} \right)\left| \phi  \right\rangle  = \left\langle \psi  \right|\left( {P\left| \phi  \right\rangle } \right),\;\;\;\;\; \left( {\left\langle \psi  \right|X} \right)\left| \phi  \right\rangle  = \left\langle \psi  \right|\left( {X\left| \phi  \right\rangle } \right).
\end{eqnarray}
So, it is natural that its corresponding scalar product could be expressed as follows
\begin{equation}\label{eq:13}
\;\;\;\;\;\;\;\;\;\left\langle {\psi }
 \mathrel{\left | {\vphantom {\psi  \phi }}
 \right. \kern-\nulldelimiterspace}
 {\phi } \right\rangle {\rm{ = }}\int_{ - {1 \mathord{\left/
 {\vphantom {1 {\sqrt \beta  }}} \right.
 \kern-\nulldelimiterspace} {\sqrt \beta  }}}^{{1 \mathord{\left/
 {\vphantom {1 {\sqrt \beta  }}} \right.
 \kern-\nulldelimiterspace} {\sqrt \beta  }}} {dp\frac{1}{{\sqrt {1 - \beta {p^2}} }}{\psi ^*}\left( p \right)\phi \left( p \right)}.
\end{equation}
Obviously, the symmetry of momentum operator is unquestionable. In fact, the position operator also has symmetry, which could be expressed by a partial integration
\begin{eqnarray}\label{eq:14}
\int_{ - {1 \mathord{\left/
 {\vphantom {1 {\sqrt \beta  }}} \right.
 \kern-\nulldelimiterspace} {\sqrt \beta  }}}^{{1 \mathord{\left/
 {\vphantom {1 {\sqrt \beta  }}} \right.
 \kern-\nulldelimiterspace} {\sqrt \beta  }}} {\frac{{dp}}{{\sqrt {1 - \beta {p^2}} }}} \psi^{*}\left( p \right)i\hbar \sqrt {1 - \beta {p^2}} \frac{d }{{d p}}\phi \left( p \right)=\int_{ - {1 \mathord{\left/
 {\vphantom {1 {\sqrt \beta  }}} \right.
 \kern-\nulldelimiterspace} {\sqrt \beta  }}}^{{1 \mathord{\left/
 {\vphantom {1 {\sqrt \beta  }}} \right.
 \kern-\nulldelimiterspace} {\sqrt \beta  }}} {\frac{{dp}}{{\sqrt {1 - \beta {p^2}} }}} {\left[ {i\hbar \sqrt {1 - \beta {p^2}} \frac{d }{{d p}}\psi \left( p \right)} \right]^*}\phi \left( p \right).
\end{eqnarray}
And the corresponding unit operator can be written as following
\begin{equation}\label{eq:15}
\;\;\;\;\;\;\;\;\;\;\;\; \int_{{{ - 1} \mathord{\left/
 {\vphantom {{ - 1} {\sqrt \beta  }}} \right.
 \kern-\nulldelimiterspace} {\sqrt \beta  }}}^{{1 \mathord{\left/
 {\vphantom {1 {\sqrt \beta  }}} \right.
 \kern-\nulldelimiterspace} {\sqrt \beta  }}} {dp} \frac{1}{{\sqrt {1 - \beta {p^2}} }}\left| p \right\rangle \left\langle p \right| = 1.
\end{equation}
Thus, the scalar product of momentum eigenstates could also be given
\begin{equation}\label{eq:16}
\;\;\;\;\;\;\;\;\;\;\;\; \left\langle {p} \mathrel{\left | {\vphantom {p {p'}}} \right. \kern-\nulldelimiterspace} {{p'}} \right\rangle  = \sqrt {1 - \beta {p^2}} \delta \left( {p - p'} \right).
\end{equation}
According to the above discussion, one obvious phenomenon is that unit operator and scalar product will vary with the structure of GUP. From the physical point of view, the modification for Heisenberg uncertainty principle with good mathematical motivation and non-perturbative structure may reveal more interesting physical phenomena, and it may be also bring new inspiration to the Hilbert space representation of quantum mechanics.
\section{Functional analysis of the position operator}
The eigenvalue problem of the position operator for the new form of GUP in momentum space can be given as
\begin{equation}\label{eq:17}
\;\;\;\;\;\;\;\;\;\;\;\; i\hbar \sqrt {1 - \beta {p^2}} \frac{d }{{d p}}{\psi _\lambda }\left( p \right) = \lambda {\psi _\lambda }\left( p \right).
\end{equation}
Thus, through some simple mathematical calculations, the expression of position eigenfunction is written as follows
\begin{equation}\label{eq:18}
\;\;\;\;\; {\psi _\lambda }\left( p \right) = {\sqrt\frac{\sqrt\beta} {\pi}}\exp \left[ { - i\frac{\lambda }{{\hbar \sqrt \beta}}\arcsin \left( {\sqrt \beta  p} \right)} \right].
\end{equation}
Therefore, the scalar product of the position eigenfunction could be given by using Eq. (\ref{eq:18})
\begin{eqnarray}\label{eq:19}
\left\langle\psi_{\lambda}(p) \mid \psi_{\lambda^{\prime}}(p)\right\rangle &=& \frac{\sqrt{\beta}}{\pi} \int_{-1 / \sqrt{\beta}}^{1 / \sqrt{\beta}} \frac{d p}{\sqrt{1-\beta p^{2}}} \exp \left[i \frac{\lambda-\lambda^{\prime}}{\hbar \sqrt{\beta}} \arcsin (\sqrt{\beta} p)\right] \nonumber \\
&=& \frac{2 \hbar \sqrt{\beta}}{\left(\lambda-\lambda^{\prime}\right) \pi} \sin \left(\frac{\lambda-\lambda^{\prime}}{2 \hbar \sqrt{\beta}} \pi\right).
\end{eqnarray}
Here, it is obvious that although the position eigenfunctions are not mutually orthogonal in general, they are still normalizable in essence. Therefore, the single parameter family of diagonalizations for the position operator could be realized by using the Eq. (\ref{eq:19}). So, the set of parameterized eigenfunctions can be written as
\begin{eqnarray}\label{eq:20}
\;\;\;\;\;\;\;\; \left\{ {\left. {\left| {{\psi _{\left( {2n + \lambda } \right)\hbar \sqrt \beta  }}} \right\rangle } \right|\;n \in Z,\;\lambda  \in \left[ { - 1,1} \right]} \right\},
\end{eqnarray}
in which any eigenfunction of the eigenfunctions family will satisfy orthogonal each other
\begin{eqnarray}\label{eq:21}
\;\;\;\;\;\;\;\; \left\langle {{{\psi _{\left( {2n + \lambda } \right)\hbar \sqrt \beta  }}}} \mathrel{\left | {\vphantom {{{\psi _{\left( {2n + \lambda } \right)\hbar \sqrt \beta  }}} {{\psi _{\left( {2n' + \lambda } \right)\hbar \sqrt \beta  }}}}}  \right. \kern-\nulldelimiterspace}
 {{{\psi _{\left( {2n' + \lambda } \right)\hbar \sqrt \beta  }}}} \right\rangle  = {\delta _{n,n'}}.
\end{eqnarray}
The Eq. (\ref{eq:21}) shows that this set is completely orthogonal normalized, and the lattice spacing of any two adjacent eigenvector is $2\hbar\sqrt\beta$,  that is to say, the distance of adjacent eigenvector is approximately equal to twice the minimal observable uncertainty of the position, that is $2\Delta X_{min}$. In addition, the expectation value of the kinetic energy operator ${P^2}/{2m}$ for these eigenvectors $|{\psi _\lambda}\rangle$ could be given as
\begin{eqnarray}\label{eq:22}
\;\;\;\;\;\;\;\; \left\langle {{\psi _\lambda }} \right|{{{P^2}} \mathord{\left/{\vphantom {{{P^2}} {2m}}} \right.  \kern-\nulldelimiterspace} {2m}}\left| {{\psi _\lambda }} \right\rangle  = \frac{1}{4m\beta}.
\end{eqnarray}
Here, we will see that the expected value of the kinetic energy operator in this paper is a finite, not an infinite like in \cite{ref13}, that is to say, the formal position eigenvectors herein can be considered as the corresponding physical states.
\begin{figure}
\includegraphics[width=0.50\textwidth]{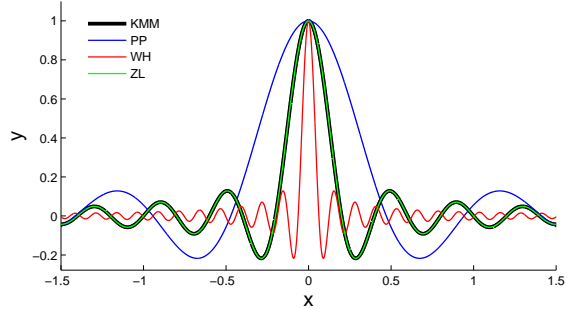}
\caption{Plotting $y=\langle \psi_{\lambda}(p)\mid \psi_{\lambda^{\prime}}(p)\rangle$ versus $x=\lambda-\lambda^{\prime}$ for Ref. \cite{ref13} (black), Ref. \cite{ref22} (blue), Ref. \cite{ref30} (red) and new GUP (green). }
\label{fig:1}
\end{figure}

Now, the scalar product of the position eigenfunctions will be further discussed herein. The Fig. 1 was drawn to describe the scalar product $\rm{y}=\langle{\psi_\lambda}\left( p \right)\mid {\psi_{\lambda^{\prime}}}\left( p \right)\rangle$ versus $x=\lambda  -\lambda^{\prime}$ and to compare it with the conclusions of other literatures. In this part of work, by comparing the Eq. (\ref{eq:19}) of this paper with the Eq. (23) given by KMM in their seminal paper \cite{ref13}, we will also see that the scalar product form of the position operator in the two articles is almost completely identical, although using the different forms of GUP. The most intuitive expression of this conclusion is that the black curve and the green curve completely overlap in Fig. 1. In addition, although the generalized uncertainty principle has been modified with different algebraic structures, the conclusions of scalar product for the four different modified models are very similar. The only difference is that the oscillating amplitude of scalar product is the most obvious in Ref. \cite{ref30}, and that in Ref. \cite{ref22} is the weakest, while the amplitude of scalar product oscillation in this paper and Ref. \cite{ref13} is between them.
\section{Maximal localization states}
In this section, let's discuss the maximal localization states in detail. Firstly, the maximal localization states satisfies the following properties
\begin{equation}\label{eq:23}
\;\;\;\; \left\langle {\psi _\xi ^{ml}} \right|X\left| {\psi _\xi ^{ml}} \right\rangle  = \xi,\;\rm{and}\;
{\left( {\Delta X} \right)_{\left| {\psi _\xi ^{ml}} \right\rangle }} = {\left( {\Delta X} \right)_{\min }}.
\end{equation}
As we have seen from the preceding discussion, the minimal uncertainty of the position can take only in the case of $\langle p \rangle=0$. Now let us give a standard derivation of the uncertain relations \cite{ref13}. For each state in the representation of Heisenberg algebra, we derive the following relationship
\begin{eqnarray}\label{eq:24}
\;\;\;\; \left| {\left( {X - \left\langle X \right\rangle  + \frac{{\left\langle {\left[ {X,P} \right]} \right\rangle }}{{2{{\left( {\Delta p} \right)}^2}}}\left( {P - \left\langle P \right\rangle } \right)} \right)\left| \psi  \right\rangle } \right| \ge 0,
\end{eqnarray}
by using Eq. (\ref{eq:24}), it can be naturally given the following relationship
\begin{eqnarray}\label{eq:25}
\;\; \left\langle \psi  \right|{\left( {X - \left\langle X \right\rangle } \right)^2} - \frac{{{{\left| {\left\langle {\left[ {X,P} \right]} \right\rangle } \right|}^{\rm{2}}}}}{{{\rm{4}}{{\left( {\Delta p} \right)}^{\rm{4}}}}}{\left( {P - \left\langle P \right\rangle } \right)^2}\left| \psi  \right\rangle  \ge 0,
\end{eqnarray}
which indicates that
\begin{eqnarray}\label{eq:26}
\;\;\;\;\;\;\;\; \Delta x\Delta p \ge \frac{{\left| {\left\langle {\left[ {X,P} \right]} \right\rangle } \right|}}{2}.
\end{eqnarray}
Therefore, it is obvious that there exists a state satisfy $\Delta x\Delta p= |\langle\left[ X,P\right]\rangle|/2$, so the maximal localization states requires this equation to satisfy
\begin{eqnarray}\label{eq:27}
\;\;\;\;\;\;\;\; \left( {X - \left\langle X \right\rangle  + \frac{{\left\langle {\left[ {X,P} \right]} \right\rangle }}{{2{{\left( {\Delta p} \right)}^2}}}\left( {P - \left\langle P \right\rangle } \right)} \right)\left| \psi  \right\rangle  = 0.
\end{eqnarray}
Substituting the coordinate and momentum operators into Eq. (\ref{eq:27}) and expanding them in the momentum representation
\begin{eqnarray}\label{eq:28}
\;\;\;\;\;\;\left[ {i\hbar \sqrt {1 - \beta {p^2}} \frac{d}{{dp}} - \left\langle X \right\rangle } \right. \left. { + \frac{{i\hbar \kappa }}{{2{{\left( {\Delta p} \right)}^2}}}\left( {\frac{p}{{\sqrt {1 - \beta {p^2}} }} - \left\langle P \right\rangle } \right)} \right]\left| \psi  \right\rangle  = 0,
\end{eqnarray}
where the parameter satisfy $\kappa {\rm{ = }}1 + \beta {\left( {\Delta p} \right)^2} + \beta {\left\langle p \right\rangle ^2}$, and the solution could be read as follows
\begin{eqnarray}\label{eq:29}
\psi \left( p \right) = N{\left( {1 - \beta {p^2}} \right)^{{\kappa  \mathord{\left/ {\vphantom {\kappa  {4\beta {{\left( {\Delta p} \right)}^2}}}} \right. \kern-\nulldelimiterspace} {4\beta {{\left( {\Delta p} \right)}^2}}}}} \exp \left[ {\left( {\frac{{\left\langle X \right\rangle }}{{i\hbar \sqrt \beta  }} + \frac{{\kappa \left\langle P \right\rangle }}{{2{{\left( {\Delta p} \right)}^2}\sqrt \beta  }}} \right)\arcsin \left( {\sqrt \beta  p} \right)} \right],
\end{eqnarray}
where $N$ is the normalization parameter. We all know that the maximal localization states can only be achieved in the case of $\langle p\rangle=0$. However, this is not a necessary and sufficient condition. So, in order to give the absolutely maximal localization states, the critical momentum uncertainty is required to reach $\Delta p= 1/\sqrt{3\beta}$. Thus, these states could be written as
\begin{eqnarray}\label{eq:30}
\;\;\;\;\psi _\xi ^{ml}\left( p \right) = \sqrt {\frac{{8\sqrt \beta  }}{{3\pi }}} \left( {1 - \beta {p^2}} \right)\exp \left( { - \frac{{i\left\langle X \right\rangle \arcsin \left( {\sqrt \beta  p} \right)}}{{\hbar \sqrt \beta  }}} \right).
\end{eqnarray}
Further using Eq. (\ref{eq:23}), we can find that the maximally localized of the states of wave function $\psi _\xi ^{ml}(p)$ will appear near $\xi$ of the momentum representation, so the Eq. (\ref{eq:30}) could be transformed into
\begin{eqnarray}\label{eq:31}
\;\;\;\;\psi _\xi ^{ml}\left( p \right) = \sqrt {\frac{{8\sqrt \beta  }}{{3\pi }}} \left( {1 - \beta {p^2}} \right)\exp \left( { - \frac{{i\xi \arcsin \left( {\sqrt \beta  p} \right)}}{{\hbar \sqrt \beta  }}} \right).
\end{eqnarray}
Here, it can be found that although these maximal localization states are not orthogonal, they still could be diagonalized. The expectation value of the kinetic energy operator $P^2/2m$ in these states of $\mid \psi _\xi ^{ml}\rangle$ can be given as following
\begin{eqnarray}\label{eq:32}
\;\;\;\;\;\;\;\; \left\langle {\psi _\xi ^{ml}} \right|\frac{{{P^2}}}{{2m}}\left| {\psi _\xi ^{ml}} \right\rangle  = \frac{1}{{12m\beta }}.
\end{eqnarray}
Similarly, the corresponding expectation values of the kinetic energy operator in the maximal localization states for different frameworks of GUP could also be given. So, the expectation values in the Ref. \cite{ref13} (KMM), Ref. \cite{ref22,ref31} (PP) and Ref. \cite{ref30} (WH) as following
\begin{eqnarray}\label{eq:33}
\;\;\;\;\;\;\;\; && \left\langle {\psi _\xi ^{ml}} \right|\frac{{{P^2}}}{{2m}}{\left| {\psi _\xi ^{ml}} \right\rangle ^{{\rm{KMM}}}} = \frac{6}{{12m\beta }},\nonumber \\
\;\;\;\;\;\;\;\; && \left\langle {\psi _\xi ^{ml}} \right|\frac{{{P^2}}}{{2m}}{\left| {\psi _\xi ^{ml}} \right\rangle ^{{\rm{PP}}}} = \frac{{0.8814}}{{12m\beta }},\\
\;\;\;\;\;\;\;\; && \left\langle {\psi _\xi ^{ml}} \right|\frac{{{P^2}}}{{2m}}{\left| {\psi _\xi ^{ml}} \right\rangle ^{{\rm{WH}}}} = \frac{{0.7176}}{{12m\beta }}.\nonumber
\end{eqnarray}
It is worth noting that although the GUP forms with different structures are adopted in this paper and in other references \cite{ref13,ref22,ref30}, the expectation value of the kinetic energy operator in the maximal localization states basically fluctuates in a very small region. By comparing the expectation values given in Eqs. (\ref{eq:32}) and (\ref{eq:33}), it can be seen that the conclusion of this paper is basically comparable to the results of references \cite{ref22,ref30}, even if compared with the results of reference \cite{ref13}, it is also only six times different. Through the comparison of the above conclusions, we speculate that the conclusion may be affected by curvature. In addition, since we are trying to introduce curvature into the new GUP in this article by modifying the position and the momentum operators at the same time, instead of modifying only one of the position or momentum operators like other authors \cite{ref13,ref19,ref30,ref31}, which seems more agree with the features of curved space-time. Therefore, from this perspective, the work of this paper may be more in line with the physical essence.
\section{A discussion for linear harmonic oscillator under the new GUP}
In this section, we will apply the proposed new form of operators to discuss the linear harmonic oscillator and further give the corresponding analytical solution. As we all know, the Hamiltonian operator of the harmonic oscillator as
\begin{eqnarray}\label{eq:34}
\;\;\;\;\;\;\;\; H = \frac{{{P^2}}}{{2m}} + \frac{{m{\omega ^2}{X^2}}}{2}.
\end{eqnarray}
Thus, considering the new form of momentum and position operators given in Eq. (\ref{eq:11}), the schrodinger equation with harmonic oscillator can be written as
\begin{eqnarray}\label{eq:35}
\;\;\;\; \left\{ {\left( {1 - \beta {p^2}} \right)\frac{{{d^2}}}{{d{p^2}}} - \beta p\frac{d}{{dp}}} \right.\left. { + \frac{{2E}}{{m{\omega ^2}{\hbar ^2}}} - \frac{{{p^2}}}{{{m^2}{\omega ^2}{\hbar ^2}\left( {1 - \beta {p^2}} \right)}}} \right\}\psi \left( p \right) = 0.
\end{eqnarray}
Next, it is easy to see that differential Eq. (\ref{eq:35}) could be transformed into the form of Eq. (\ref{eq:36}) by making a variable $t=(1-\sqrt \beta p)/2$
\begin{eqnarray}\label{eq:36}
\;\;\;\;\left[ {\left( {t - {t^2}} \right)\frac{{{d^2}}}{{d{t^2}}} + \left( {\frac{1}{2} - t} \right)\frac{d}{{dt}} + \frac{{2E}}{{\beta m{\omega ^2}{\hbar ^2}}}} \right. \left. { - \frac{1}{{4{\beta ^2}{m^2}{\omega ^2}{\hbar ^2}}}\left( {\frac{1}{t} - 4 + \frac{1}{{1 - t}}} \right)} \right]\psi \left( t \right) = 0.
\end{eqnarray}
It is exciting that the Eq. (36) could transform into an equation by extracting the appropriate asymptotic behavior, which is formally consistent with the standard equation of the Bethe ansatz method \cite{ref33,ref34,ref35,ref36,ref37,ref38,ref39,ref40}, and the corresponding exact solution can be derived. In other words, the wave function needs to be transformed as follows
\begin{equation}\label{eq:37}
\;\;\;\;\;\;\;\;\;\;\psi \left( t \right) = {\left( {t - {t^2}} \right)^\alpha }f\left( t \right),
\end{equation}
in which $\alpha$ is a parameter related to $\beta$. Bring the Eq. (\ref{eq:37}) into Eq. (\ref{eq:36}), it can be find that the equation requires the parameter meeting
\begin{eqnarray}\label{eq:38}
\;\;\;\;\;\;\;\;\;\; \alpha  = \frac{1}{4} + \frac{{\sqrt {{\beta ^2}{m^2}{\omega ^2}{\hbar ^2} + 4} }}{{4\beta m\omega \hbar }}.
\end{eqnarray}
Therefore, the corresponding differential equation could be written as
\begin{eqnarray}\label{eq:39}
\;\;\;\;\;  \left( {t - {t^2}} \right)f''\left( t \right) + \left[ {\frac{1}{2}} + 2\alpha - \left( {1 + 4\alpha } \right)t \right]f'\left( t \right) + \left( {\frac{{2\beta mE + 1}}{{{\beta^2} m^2 {\omega ^2}{\hbar ^2}}} - 4{\alpha ^2}} \right)f\left( t \right) = 0.
\end{eqnarray}
So, the corresponding degree $n$ polynomial solutions of the differential Eq. (\ref{eq:39}) read
\begin{equation}\label{eq:40}
\;\;\;\;\;\;\; f\left( t \right) = \prod\limits_{i = 1}^n {\left( {t - {t_i}} \right)},\;\;\;f\left( t \right) \equiv 1\;\;\;\rm{for}\;\;\;n = 0,
\end{equation}
with different roots $\{t_{i}\}$. Moreover, the corresponding eigenfunction and eigenvalue of the linear harmonic oscillator in the context of new GUP could be expressed as
\begin{eqnarray}\label{eq:41}
\;\;\;\;\;\;\; {E_n} = \left( {n + \frac{1}{2}} \right)\hbar \omega \sqrt {1 + \frac{{{\beta ^2}{m^2}{\omega ^2}{\hbar ^2}}}{4}} + \left( {{n^2} + n + \frac{1}{2}} \right)\frac{{\beta m{\omega ^2}{\hbar ^2}}}{2},
\end{eqnarray}
\begin{eqnarray}\label{eq:42}
\;\;\;\;{\psi _n}\left( p \right) = {\left( {\frac{1}{4} - \frac{{\beta {p^2}}}{4}} \right)^\alpha }\left[ {\prod\limits_{i = 1}^n {\left( {\frac{1}{2} - \frac{{\sqrt \beta  }}{2}p - {t_i}} \right)} } \right].
\end{eqnarray}
Here, the roots $\{t_{i}\}$ can be determined according to Bethe ansatz equations
\begin{eqnarray}\label{eq:43}
\;\;\;\; \sum\limits_{j \ne i}^n {\frac{2}{{{t_i} - {t_j}}}}+ \frac{{1 + 4\alpha  - \left( {2 + 8\alpha } \right){t_i}}}{{2\left( {{t_i} + t_i^2} \right)}} = 0,\;\;\;\;i = 1,2,...,n.
\end{eqnarray}
Next, the energy spectrum and wave function of the ground state and the first excited state system as the special cases of general expression can be derived from Eqs. (\ref{eq:41}) and (\ref{eq:42}). And solutions of the ground state corresponds to the case where $n=0$
\begin{eqnarray}\label{eq:44}
\;\;\;\;\; &&{\psi _0}\left( p \right) = {\left( {\frac{1}{4} - \frac{{\beta {p^2}}}{4}} \right)^\alpha },\nonumber \\
&& {E_0} = \frac{1}{2}\hbar \omega \sqrt {1 + \frac{{{\beta ^2}{m^2}{\omega ^2}{\hbar ^2}}}{4}}  + \frac{{\beta m{\omega ^2}{\hbar ^2}}}{4}.
\end{eqnarray}
The solution of first excited state corresponds to the general expression of the case $n=1$, the corresponding analytic expression could also be written as
\begin{eqnarray}\label{eq:45}
\;\;\;\;\; && {\psi _1}\left( p \right) = {\left( {\frac{1}{4} - \frac{{\beta {p^2}}}{4}} \right)^\alpha }\left( {\frac{1}{2} - \frac{{\sqrt \beta  }}{2}p - {t_1}} \right), \nonumber\\
&& {E_1} = \frac{3}{2}\hbar \omega \sqrt {1 + \frac{{{\beta ^2}{m^2}{\omega ^2}{\hbar ^2}}}{4}}  + \frac{{5\beta m{\omega ^2}{\hbar ^2}}}{4},
\end{eqnarray}
in which the root $t_{1}$ is calculated by Bethe ansatz equation
\begin{eqnarray*}
\;\;\;\;\;\;\;\; 1 + 4\alpha  - \left( {2 + 8\alpha } \right){t_1} = 0\;\;\; \Rightarrow \;\;\;{t_1} = {1 \mathord{\left/
 {\vphantom {1 2}} \right.  \kern-\nulldelimiterspace} 2}.
\end{eqnarray*}
It can be found that as the parameter $\beta$ in Eq. (\ref{eq:35}) approaching zero, not only the Schrodinger equation with linear harmonic oscillator in ordinary quantum mechanics is completely reproduced, but also the corresponding analytical solution could also be directly degenerated from the exact solution of the new GUP. Obviously, the corresponding energy spectrum of ordinary quantum mechanics could be expressed as
\begin{eqnarray}\label{eq:46}
\;\;\;\;\;\;\;\; E_n=(n+\frac{1}{2})\hbar \omega.
\end{eqnarray}
Here, it is easy to verify that the conclusion of generalized uncertainty principle in limit $\beta \to 0$ can well reproduce the conclusion of ordinary quantum mechanics, which implying that the ordinary quantum mechanics can be regarded as a special case of the new form of GUP at $\beta=0$. It is not difficult to understand that because $\beta$ is related to curvature, when its value is zero, it indicates that the curvature of space-time is zero at this time, and the space-time studied by ordinary quantum mechanics is also a space-time with zero curvature. Thus, this phenomenon also provides a strong support for $\beta$ is regarded as a curvature related parameter, and further verifies the correctness of the new form of GUP. By comparing the Eqs. (\ref{eq:41}) and (\ref{eq:46}), for a finite $\beta$, the energy spectrum of GUP shows that it is not only related to the principal quantum number $n$, but also to the square of $n$. In the low energy region, the difference between any two energy spectra for different $\beta$ is small, but as the principal quantum number increases, the increasing speed of the gap between any two energy spectra becomes faster and faster. Especially when $n$ increases to a certain degree, the variation of energy spectrum will mainly depend on the square of $n$. Similarly, we can also see that for a given energy level, the energy of the harmonic oscillator will increase as $\beta$ increases. So, in order to show this phenomenon more vividly, the picture of harmonic oscillator energy spectrum with the change of principal quantum number $n$ is plotted in Fig. 2.
\begin{figure}
\includegraphics[width=0.50\textwidth]{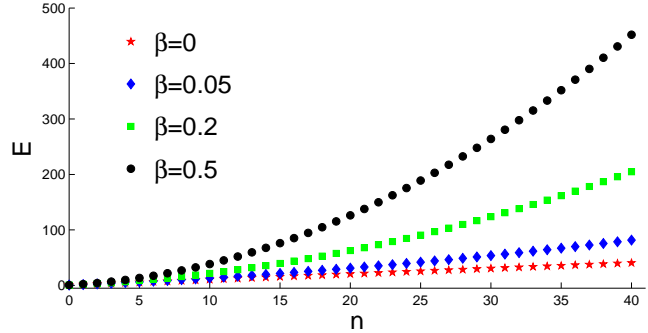}
\caption{Fig.2 The energy spectrum versus $n$ for the different value $\beta$.}
\label{fig:2}
\end{figure}
\section{Conclusion}
In this paper, a new form of higher-order GUP is proposed by constructing the new operators, which makes the modified Heisenberg uncertainty principle seem more in line with the feature of curved space-time. Through further research, it is also found that the operator expressions are given in the momentum representation of this paper not only perfectly follows the previous conclusions, but also has a good performance in the process of dealing with problems. Then, the new GUP is used in subsequent research to further discuss the maximum localization states and position eigenfunctions. By comparing the above conclusions with the existing literature, it can be found that the new revised GUP proposed by us is further supported. Finally, the harmonic oscillator problem, as an example, is studied in detail, and it can be seen that the new form of operator in this part of research shows incomparable superiority in dealing with problems. Moreover, the exact solutions of the harmonic oscillator in ordinary quantum mechanics can be perfectly reproduced by the analytical solutions of the new GUP, which undoubtedly further enhances the persuasiveness for our work.
\section{Acknowledgements}
This paper is supported by the National Natural Science Foundation of China (Grant Nos: 11465006 and 11565009) and the Major Research Project of innovative Group of Guizhou province (2018-013).



\end{document}